\documentclass[sigconf]{acmart}

\setcopyright{none}
\renewcommand\footnotetextcopyrightpermission[1]{}

\usepackage{amsmath}
\usepackage{algorithm}
\usepackage{algpseudocode}
\usepackage{booktabs}
\usepackage{graphicx}
\usepackage{listings}
\usepackage{microtype}
\usepackage{xcolor}

\newcommand{\enabled}{\mathsf{enabled}}
\newcommand{\frontier}{\mathsf{frontier}}
\newcommand{\oracle}{\mathcal{O}}
\newcommand{\fidelity}{\mathsf{Fid}}
\newcommand{\depthscore}{\mathsf{Depth}}
\newcommand{\linset}{\mathsf{Lin}}
\newcommand{\execset}{\mathsf{Exec}}

\lstdefinestyle{ruststyle}{
  basicstyle=\ttfamily\scriptsize,
  keywordstyle=\bfseries,
  commentstyle=\itshape,
  columns=fullflexible,
  keepspaces=true,
  showstringspaces=false,
  frame=single,
  aboveskip=4pt,
  belowskip=4pt
}

\begin{document}

\title{From Resource Flow to Executable Tests: Petri-Net-Guided LLM Test Generation for Concurrent Stateful Rust APIs}

\author{Kaiwen Zhang, Guanjun Liu}
\affiliation{%
  \institution{Tongji University}
  \country{Shanghai,China}
}
\email{zhangkw@tongji.edu.cn}

\begin{abstract}
Concurrent stateful library APIs expose behavior through evolving resource ownership, lifecycle states, and competing interleavings. Large language models can synthesize executable Rust tests, but their outputs often violate API preconditions, remain shallow, or reduce concurrency to accidental sequential traces. Conversely, model-based and systematic testing techniques provide semantic control but commonly require substantial handwritten code to turn abstract scenarios into executable tests. This paper addresses the gap between formal scenario design and low-cost test concretization.

We present a Petri-net-guided methodology for test generation over concurrent stateful Rust APIs. The method represents API resources, lifecycle conditions, and causal dependencies as colored tokens and transitions; derives legal deep-state, near-legal, and partial-order concurrent scenarios; and uses these scenarios as a constrained intermediate representation for LLM-based code synthesis. A local-faithfulness contract and structural repair loop preserve the modeled intent during concretization, while Petri-guided schedule shaping prioritizes high-conflict concurrency skeletons for systematic exploration. A layered semantic oracle then distinguishes synthesis failures from violations of the target API's expected behavior.

We implement a prototype for Rust concurrency libraries and define an evaluation protocol that examines executability, structural fidelity, deep-state reachability, boundary-failure yield, and conflict coverage. The central methodological principle is to separate responsibilities: the Petri net specifies semantic intent, resource flow, reachability, conflict, and bug-relevant mutations, whereas the LLM realizes library-specific syntax, task scaffolding, and assertions. This design provides a concrete basis for studying whether formal resource-flow models can make LLM-generated concurrency tests more faithful, stateful, and diagnostically useful.
\end{abstract}



\maketitle

\section{Introduction}
Many Rust library APIs are not isolated function interfaces. They expose handles, permits, closures, buffered state, and task-level interaction patterns whose behavior depends on prior operations and competing schedules. In this setting, Rust's type system removes broad classes of memory errors, but it does not prove that a library respects its higher-level protocol. The bugs that remain are semantic faults: a stale capability is still accepted, a close operation invalidates the wrong behavior, a buffered value is lost, or a test hangs only after a specific interleaving reaches a deep state.

These APIs are difficult to test for three coupled reasons. First, interesting behaviors are strongly stateful: many failure modes appear only after a legal prefix establishes a nontrivial resource configuration. Second, the prefix itself must be semantically disciplined; otherwise later observations are meaningless because the test never exercised a valid protocol state. Third, concurrency matters at the level of partial order rather than raw length. A long sequence is not useful if it accidentally serializes the very race that should have been exposed.

Existing approaches each cover only part of this space. Model-based and dependency-aware testing can describe legal states, resource flow, and boundary conditions, but converting those abstractions into executable Rust tests still demands substantial handwritten scaffolding, task orchestration, and API-specific assertions. Direct LLM prompting reduces that coding burden, but it also pushes semantic responsibility back into the model. In practice, the LLM may invent events, violate enablement conditions, confuse reserved capabilities with fresh ones, weaken assertions, or flatten concurrent behavior into a convenient sequential trace.

Our position in this paper is intentionally narrower. We do not ask the LLM to discover concurrency semantics from scratch, and we do not claim to eliminate modeling effort. Instead, we use a Petri-net model to encode resource multiplicity, causality, conflict, and near-legal boundary cases, then use the LLM only to realize model-produced scenarios as executable Rust code. The scheduler still explores runtime interleavings, and API-specific invariants still need to be authored. The point is to separate responsibilities so that semantic intent stays outside the LLM while code realization stays inexpensive.

This tradeoff fits concurrent stateful Rust APIs especially well. A Petri net can represent senders, receivers, permits, capacity, buffered messages, and observation obligations as tokens, and it can express the precise conflicts and causal dependencies that make a test diagnostically useful. Once that structure is explicit, the LLM can do what it is relatively good at: writing target-specific setup, task creation, assertions, and cleanup. The key design rule is simple: the model writes meaning, the LLM writes code.

The price of this design is also clear. SyncPetri requires manually authored resource models, adapters, and scenario invariants, and its schedule shaping improves harness selection above tools such as Loom rather than replacing their internal search. We state these limitations up front because they define the appropriate contribution: not fully automatic concurrency verification, but a testing architecture that trades some modeling effort for stronger semantic control over generated tests.

Concretely, SyncPetri synthesizes legal deep-state traces, near-legal boundary probes, and partial-order concurrent scenarios from a colored Petri net; concretizes them through a constrained prompt and structural repair loop; and judges the resulting executions with a layered oracle that separates concretization failure from semantic failure. The result is a testing pipeline designed for concurrent Rust APIs whose hard bugs live in resource protocols rather than in isolated function outputs.

We make five contributions:
\begin{itemize}
  \item We formalize concurrent stateful Rust API testing as Petri-net-guided scenario synthesis over resource-carrying transitions.
  \item We define an adapter schema and a local-faithfulness contract that connects abstract Petri-net steps to concrete Rust executions.
  \item We define a scenario representation that unifies legal reachability, near-legal boundary mutation, and partial-order concurrency.
  \item We present a constrained concretization loop and a multi-layer oracle that distinguish generation failure from semantic failure.
  \item We propose a Petri-guided schedule shaping method for Loom-compatible harnesses and instantiate the full workflow on \texttt{tokio::sync}-style APIs.
\end{itemize}

\section{Problem Setting and Running Example}
\label{sec:problem}
We target APIs with three fundamental properties: 1) Statefulness: The outcome or validity of an operation depends heavily on the internal abstract resource state of the system. 2) Resource sensitivity: Operations directly manipulate capabilities—consuming, producing, splitting, merging, or invalidating handles, permits, buffered data, or access tokens. 3) Concurrency exposure: Multiple distinct tasks or handles interact simultaneously, meaning distinct execution schedules can yield entirely different observable outcomes.

This covers a useful slice of Rust libraries. In \texttt{tokio::sync}, for example, \texttt{mpsc} channels expose senders, receivers, permits, buffering, and closure; \texttt{watch} exposes latest-value semantics and observed states; \texttt{broadcast} exposes lagging receivers and replay boundaries; and \texttt{Semaphore} exposes acquisition, release, and closure behavior~\cite{tokio_mpsc_docs,tokio_watch_docs,tokio_broadcast_docs,tokio_semaphore_docs}. These APIs are small enough to test locally but rich enough to require genuine semantic orchestration. Table~\ref{tab:mpscmodel} shows a compact abstraction for a running \texttt{mpsc} example.

\begin{table}[t]
\caption{Example Petri-net places for a bounded \texttt{mpsc}.}
\label{tab:mpscmodel}
\centering
\small
\begin{tabular}{p{0.28\columnwidth}p{0.62\columnwidth}}
\toprule
Place & Meaning \\
\midrule
\texttt{LiveSender(s)} & Sender handle $s$ exists and may initiate send-side operations. \\
\texttt{LiveReceiver(r)} & Receiver handle $r$ exists and may receive or close. \\
\texttt{Open(c)} & Channel $c$ accepts ordinary sends. \\
\texttt{Closed(c)} & Channel $c$ has been closed from the receive side. \\
\texttt{Permit(s,c)} & Sender $s$ holds a reserved permit on channel $c$. \\
\texttt{Cap(c,n)} & Channel $c$ has $n$ unreserved buffer slots. \\
\texttt{Buf(c,n)} & Abstract buffered-item count for channel $c$. \\
\texttt{Obs(x)} & Runtime observation token emitted by instrumentation. \\
\bottomrule
\end{tabular}
\end{table}
The research problem is therefore: \textit{How can we synthesize semantically meaningful concurrent test scenarios from a Petri-net model of a Rust API, concretize those scenarios into executable Rust code with an LLM, and judge whether the resulting execution preserves or violates the intended API semantics?}

\subsection{Bug Model}
We target semantic faults that escape Rust's type system and strict memory-safety checks. We classify these faults into four core families:
\begin{itemize}
    \item $\mathbf{B_{\text{pre}}}$ (\textbf{Precondition Enforcement}): An operation incorrectly succeeds even though its abstract prerequisite condition is violated (e.g., sending into a closed channel).
    \item $\mathbf{B_{\text{state}}}$ (\textbf{Post-State Inconsistency}): Resource accounting, ownership transfer, or cleanup logic diverges from the specification after an operation executes.
    \item $\mathbf{B_{\text{race}}}$ (\textbf{Race Sensitivity}): A schedule-dependent interleaving exposes an observation sequence or state variation that the high-level protocol strictly forbids.
    \item $\mathbf{B_{\text{live}}}$ (\textbf{Liveness and Blocking}): The implementation suffers from permanent blocking, dropped wake-ups, or a failure to terminate after reaching a terminal protocol state.
\end{itemize}

For \texttt{tokio::sync} APIs, these categories correspond to concrete fault shapes such as send-after-close success, stale-handle acceptance, lost notification after value update, inconsistent permit accounting, or non-terminating close races. This bug model serves as the ground truth for our seeded mutant evaluation.

\subsection{Running Example and Baseline Pitfalls}
\label{sec:motivation}
We motivate the SyncPetri workflow using a capacity-one Tokio MPSC channel. The setup involves a receiver $R_0$, an ordinary sender $S_p$ used to reserve capacity, and an independent sender $S_b$ acting as a boundary probe. 

The core semantic invariant of interest involves a subtle asynchronous contract: closing the receiver must immediately reject new ordinary send operations, yet any pre-existing \texttt{Permit} or \texttt{OwnedPermit} successfully acquired \textit{prior} to closure remains valid and must be allowed to commit its message to the buffer~\cite{tokio_mpsc_docs}. The receiver is then obligated to drain this outstanding message before final termination.

This nuanced lifecycle contract creates several plausible semantic faults: an implementation might erroneously drop the message backed by the outstanding permit upon closure, reject the valid permit commit, or hang indefinitely due to internal desynchronization between close flags and permit counters. 

Traditional testing paradigms struggle to isolate this behavior. A purely random concurrency fuzzer is highly unlikely to generate the exact interleaved sequence required to hit this deep state (reserving capacity, racing a close with a commit, and then validating the residual drain). Conversely, a pure prompt-only LLM routinely suffers from two opposing failures: it either collapses the concurrent execution into a simplified sequential trace that entirely misses the race, or it hallucinates a strict sequential rule (e.g., assuming \textit{all} operations fail post-close), thereby generating incorrect test assertions that mismatch the API's actual contract.

\paragraph{Step 1: Resource Model and Deep State.}
Initially, the abstract initial marking $M_0$ contains two live sender capabilities, one live receiver, an open-channel token, and one capacity token. The first modeled event explicitly targets a deep state by reserving capacity:
\[
e_1 = \mathsf{reserveOwned}(S_p) \rightarrow P_0.
\]
The resulting marking moves the net into a distinct configuration containing $\texttt{Permit}(P_0, c)$ and $\texttt{Cap}(c, 0)$. This represents the deep state of interest: the channel possesses an outstanding, isolated send capability ($P_0$) distinct from an ordinary open sender or a simple buffered item. Crucially, the model dictates that the permit-commit transition consumes $P_0$ without requiring $\texttt{Open}(c)$, while ordinary sends still depend on both $\texttt{Open}(c)$ and free capacity.

\paragraph{Step 2: Scenarios.}
From this marking, SyncPetri builds a partial-order graph over $E=\{e_1,\ldots,e_7\}$ with the following labels:
\begin{align*}
e_1 &: \mathsf{reserveOwned}(S_p) \rightarrow P_0, &
e_2 &: \mathsf{spawn}(T_1,P_0),\\
e_3 &: \mathsf{permitSend}(P_0,m_1), &
e_4 &: \mathsf{close}(R_0),\\
e_5 &: \mathsf{recv}(R_0) \rightarrow m_1, &
e_6 &: \mathsf{trySend}(S_b,m_2),\\
e_7 &: \mathsf{recvEnd}(R_0). &&
\end{align*}
Rather than fixing a rigid, linear trace, only necessary causal dependencies are captured in $\prec$:
\[
\begin{gathered}
e_1 \prec e_2 \prec e_3,\qquad e_1 \prec e_4,\\
e_3 \prec e_5,\qquad e_4 \prec e_5 \prec e_6 \prec e_7.
\end{gathered}
\]
Events $e_3$ and $e_4$ are causally independent but compete for overlapping resource fields, and are thus flagged as an explicit race pair: $(e_3,e_4)\in\#$. This partial order safely encapsulates two valid interleavings: commit-before-close and close-before-commit. Event $e_6$ acts as a near-legal boundary probe; following $e_5$, the channel's capacity is free but its state remains closed, meaning $e_6$ violates exactly one enablement condition ($\texttt{Open}(c)$). The expected observations are mapped as set-valued allowed classes:
\[
\begin{aligned}
\Phi_{e_3}&=\{\mathsf{SendOk}\}, &
\Phi_{e_5}&=\{\mathsf{Msg}(m_1)\},\\
\Phi_{e_6}&=\{\mathsf{ClosedLike}\}, &
\Phi_{e_7}&=\{\mathsf{End}\}.
\end{aligned}
\]

\begin{table}[t]
\caption{SyncPetri Responsibilities Division and Artifacts for the Motivating Example.}
\label{tab:motivation-flow}
\centering
\small
\setlength{\tabcolsep}{3pt}
\begin{tabular}{@{}p{0.15\columnwidth}p{0.36\columnwidth}p{0.38\columnwidth}@{}}
\toprule
Stage & Model intent & Realization \\
\midrule
\textbf{Scenario} & 7 typed events, causal edges $\prec$, conflict pair $(e_3, e_4)$ & Immutable JSON prompt constraints. \\
\midrule
\textbf{Code} & Resource bindings, outcome classes $\Phi$, banned edits & LLM generates tasks, scopes, adapters, and assertions. \\
\midrule
\textbf{Schedule} & Frontier prioritized by uncovered conflicts & Two shaped execution harnesses. \\
\midrule
\textbf{Oracle} & $\mathcal{O}_{\text{str}} \land \mathcal{O}_{\text{out}} \land \mathcal{O}_{\text{inv}} \land \mathcal{O}_{\text{live}}$ & Separates synthesis failures from API bugs. \\
\bottomrule
\end{tabular}
\end{table}

\paragraph{Step 3: Constrained LLM Concretization.}
The structural graph is serialized into a highly constrained data prompt rather than a loose text description. The prompt provides explicit resource mappings, edge dependencies, and allowable outcome envelopes. 

To bridge the gap between abstract events and executable tasks, the framework derives execution scaffolding from the conflict relation. Because $(e_3, e_4) \in \#$ is an active concurrency pair, the prompt requires explicit readiness, release, and completion handshakes rather than letting the model serialize the actions. Listing~\ref{lst:motivating-harness} illustrates the resulting code artifact. The LLM remains free to handle target-specific syntax and variable scoping but cannot strip the structural instrumentation markers (\texttt{mark}).

\begin{lstlisting}[style=ruststyle,caption={Schematic generated harness for the motivating scenario.},label={lst:motivating-harness}]
let (sp, sb, mut r0) = bounded_with_clone(1);
let p0 = mark(e1, || reserve_owned(sp));
let gate = deterministic_handshake();
let child = mark(e2, || spawn(move || {
    gate.ready(); gate.wait_release();
    mark(e3, || p0.send(m1))
}));
gate.wait_ready(); gate.release();
mark(e4, || r0.close());
join_bounded(child);
assert_msg(mark(e5, || recv(&mut r0)), m1);
assert_closed(mark(e6, || sb.try_send(boundary_msg)));
assert_end(mark(e7, || recv_bounded(&mut r0)));
\end{lstlisting}

\paragraph{Step 4: Petri-Guided Schedule Exploration.}
The synthesized event graph induces two macro harnesses:
\[
\begin{aligned}
h_{\mathit{commit}}&=[e_1,e_2,e_3,e_4,e_5,e_6,e_7],\\
h_{\mathit{close}}&=[e_1,e_2,e_4,e_3,e_5,e_6,e_7].
\end{aligned}
\]
The generated handshakes guarantee that the deep state $P_0$ is established before the race window opens. A deterministic scheduler adapter (or Loom) then enumerates only the micro-interleavings inside that window. This division of labor lets the Petri-net model select the macro-concurrency surface while the runtime scheduler explores local execution choices.

\paragraph{Step 5: Oracle Decision and Report.}
On a compliant implementation, either linearization may occur at runtime. Regardless of whether $e_3$ or $e_4$ executes first, $e_3$ must yield $\textsc{SendOk}$, $e_5$ must observe $m_1$, and the post-close boundary check $e_6$ must evaluate to $\textsc{ClosedLike}$. 

The layered oracle first invokes the structural checker ($\mathcal{O}_{\text{str}}$) to verify that all seven runtime markers executed in a sequence that satisfies $\prec$. If an LLM error or compiler optimization reordered or bypassed a marker, the run is flagged as a \textsc{ConcretizationError} and pruned. If $\mathcal{O}_{\text{str}}$ passes but the library returns an unmapped observation class (e.g., if a mutant rejects the outstanding permit during a close-first interleaving), the system registers a true $\textsc{SemanticFailure}$. The resulting diagnostic profile is recorded as:
\[
\mathsf{Report} = (\mathcal{S}_{pc}, \textsc{SemanticFailure}, \{\mathit{out}\}, \tau^\star, h_{\mathit{close}}).
\]
This structured output ensures that hundreds of identical low-level scheduler interleavings that expose the exact same root bug are seamlessly de-duplicated into a single actionable report.

\section{Formal Model}
\label{sec:model}

\subsection{API Abstraction and Colored Petri Net}

We begin with a lightweight API abstraction
\[
A = (R, O, \Sigma, \Omega),
\]
where $R$ is a set of resource sorts, $O$ is a set of operation names, $\Sigma$ maps each operation to typed arguments and result classes, and $\Omega$ is a set of observation classes used by the runtime oracle.

The API abstraction is compiled into a colored Petri net
\[
N_A = (P, T, F, \chi, \lambda, g, u, M_0).
\]
where $P$ is a finite set of places, $T$ is a finite set of transitions, $F \subseteq (P \times T) \cup (T \times P)$ is the flow relation, $\chi$ maps each place to a token color domain, $\lambda : T \rightarrow O$ labels transitions with API or harness operations, $g_t$ is a guard predicate for each transition $t$, $u_t$ is a token-update function for each transition $t$, $M_0$ is the initial marking.

Let $I_t(p)$ and $O_t(p)$ denote the input and output token multisets of transition $t$ at place $p$. A transition is enabled under marking $M$ iff
\[
\enabled(M,t) \iff g_t(M) = \texttt{true} \;\land\; \forall p \in P,\; I_t(p) \subseteq M(p).
\]
If $\enabled(M,t)$ holds, firing $t$ yields a new marking $M'$ defined by
\[
M' = u_t\big((M - I_t) + O_t\big).
\]
We write $M \xrightarrow{t} M'$ for one firing step and $M_0 \xrightarrow{t_1 \cdots t_k} M_k$ for a finite firing sequence. The reachability set is
\[
\mathsf{Reach}(N_A) = \{ M \mid \exists \pi,\; M_0 \xrightarrow{\pi} M \}.
\]

The practical purpose of the net is not only to reject impossible calls. It records \emph{how} resources move, \emph{which} operations compete for them, and \emph{which} events are causally independent.

\subsection{Adapter Schema and Local Faithfulness}

The Petri net is abstract, but the test oracle runs over concrete Rust executions. We therefore associate each target library domain $d$ with an adapter schema
\[
A_d = (\mathsf{Ctor}, \mathsf{Step}, \mathsf{Spawn}, \mathsf{Mark}, \mathsf{Assert}, \mathsf{Cleanup}, \Gamma, \rho_d).
\]
Here $\mathsf{Ctor}$ constructs the initial runtime objects, $\mathsf{Step}$ maps a modeled transition to a concrete Rust operation, $\mathsf{Spawn}$ packages spawnable event groups into tasks, $\mathsf{Mark}$ emits event markers, $\mathsf{Assert}$ instantiates concrete checks, $\mathsf{Cleanup}$ bounds teardown, $\Gamma$ maps concrete results to observation classes in $\Omega$, and $\rho_d$ abstracts a concrete runtime state to a Petri-net marking.

We do not require a full bisimulation between implementation and model. The methodological requirement is a \emph{local faithfulness} contract at modeled event boundaries. For an event $e$ with $\eta(e)=t$, let
\[
\mathsf{Step}_d(t,\sigma,\beta(e)) \leadsto (\sigma',o)
\]
denote one concrete adapter step from runtime state $\sigma$ to $\sigma'$ with raw observation $o$, and let $\Omega_t^{ok},\Omega_t^{err}\subseteq\Omega$ denote the modeled success and error observation classes of $t$. For legal steps, we require
\[
\begin{aligned}
\rho_d(\sigma)=M \land M \xrightarrow{t} M'
\Longrightarrow {} &
\exists \sigma',o.\;
\mathsf{Step}_d(t,\sigma,\beta(e)) \leadsto (\sigma',o) \\
& \land \rho_d(\sigma')=M'
\land \Gamma(o)\in\Omega_t^{ok}.
\end{aligned}
\]
For near-legal steps, we require that a single-precondition violation maps to an explicit error-class outcome:
\[
\begin{aligned}
\rho_d(\sigma)=M \land {} &
|\Delta_F(M,t)| + |\Delta_G(M,t)| = 1 \\
\Longrightarrow {} &
\exists \sigma',o.\; 
\mathsf{Step}_d(t,\sigma,\beta(e)) \leadsto (\sigma',o) \\
& \land \Gamma(o)\in\Omega_t^{err}.
\end{aligned}
\]
This contract is intentionally lightweight. It requires the adapter to preserve the meaning of modeled steps and boundary failures without demanding that every internal library state be exposed or reconstructed.

\subsection{Reusable Transition Schemas}

To avoid making each Petri net a one-off artifact, we model concurrent Rust APIs with a small library of reusable transition schemas. A schema is a tuple
\[
\theta =
\bigl(P^{in}_{\theta}, P^{out}_{\theta}, g_{\theta}, u_{\theta},
\Omega_{\theta}\bigr),
\]
consisting of typed input places, typed output places, a guard, a token-update rule, and the observation classes attached to the step. Instantiating a schema only requires binding symbolic place names and resource identifiers.

Four schemas are especially common in \texttt{tokio::sync} APIs:
{\small
\[
\begin{array}{@{}l@{}}
\theta_{\text{clone}}:\\
\qquad \texttt{LiveHandle}(x)
\rightarrow
\texttt{LiveHandle}(x) + \texttt{LiveHandle}(x'),\\[2pt]
\theta_{\text{reserve}}:\\
\qquad \texttt{LiveSender}(s) + \texttt{Open}(c) + \texttt{Cap}(c,n)\\
\qquad \rightarrow
\texttt{LiveSender}(s) + \texttt{Permit}(s,c) + \texttt{Cap}(c,n\!-\!1),\\[2pt]
\theta_{\text{close}}:\\
\qquad \texttt{LiveReceiver}(r) + \texttt{Open}(c)
\rightarrow
\texttt{LiveReceiver}(r) + \texttt{Closed}(c),\\[2pt]
\theta_{\text{consume}}:\\
\qquad \texttt{LiveReceiver}(r) + \texttt{Buf}(c,n)\\
\qquad \rightarrow
\texttt{LiveReceiver}(r) + \texttt{Buf}(c,n\!-\!1)\\
\qquad {} + \texttt{Obs}(\mathsf{RecvOk}).
\end{array}
\]
}
The guards on $\theta_{\text{reserve}}$ and $\theta_{\text{consume}}$ additionally require $n>0$. Near-legal mutations then arise naturally by violating exactly one of these guard or token conditions, such as attempting $\theta_{\text{reserve}}$ when the channel is closed or capacity is exhausted.

These schemas transfer across the target API family. In \texttt{mpsc}, $\theta_{\text{reserve}}$ models permit acquisition; in \texttt{Semaphore}, the same schema models acquisition over capacity tokens; in \texttt{watch}, $\theta_{\text{consume}}$ becomes observation of an unseen update; and in \texttt{broadcast}, it becomes lag-sensitive receive with a richer outcome-class mapping. The point is methodological: the Petri model is not handwritten from scratch for every call sequence, but assembled from resource-transition idioms that recur across concurrent Rust libraries.

\subsection{Scenario Semantics}

An abstract scenario is a tuple
\[
\mathcal{S} = (E, \prec, \#, \eta, \beta, \Phi),
\]
where $E$ is a finite event set, $\prec \;\subseteq E \times E$ is a strict partial order, $\# \;\subseteq E \times E$ is a symmetric conflict relation, $\eta : E \rightarrow T$ maps events to Petri-net transitions, $\beta$ binds symbolic resources and payloads to event parameters, $\Phi$ contains expected outcome predicates and global invariants.

A \emph{linearization} of $\mathcal{S}$ is any bijective sequence over $E$ that respects $\prec$. We write
\[
\linset(\mathcal{S}) =
\{ \pi \mid \pi \text{ is a linearization of } (E,\prec) \}.
\]
The executable semantics of a scenario under net $N_A$ is
\[
\execset(\mathcal{S},N_A)=
\{ \pi \in \linset(\mathcal{S}) \mid M_0 \xrightarrow{\eta(\pi)} M \text{ for some } M \},
\]
where $\eta(\pi)$ lifts $\eta$ pointwise from events to transition sequences. A scenario is \emph{legal} iff $\execset(\mathcal{S},N_A)\neq \emptyset$. Equivalently, at least one linearization $e_1,\dots,e_n$ satisfies $M_0 \xrightarrow{\eta(e_1)\cdots\eta(e_n)} M_n$.

To formalize boundary mutation, let $\Delta_F(M,t) = \{ p \in P \mid I_t(p) \nsubseteq M(p) \}$,
and assume the guard of $t$ is written as $g_t = q_1 \land q_2 \land \cdots \land q_r$. Let $\Delta_G(M,t) = \{ q_j \mid q_j(M)=\texttt{false},\; 1 \leq j \leq r \}$, so that missing tokens and violated atomic guards are counted separately. A disabled transition $t$ is \emph{near-legal} at $M$ if $|\Delta_F(M,t)| + |\Delta_G(M,t)| = 1$. This definition captures the specific kind of semantic boundary case we want: a trace that is almost enabled, but violates exactly one precondition.

The conflict relation $\#$ records pairs of events that should be scheduled adversarially because they compete for a token class, touch the same linear resource, or correspond to user-declared race pairs. The partial order $\prec$ records only necessary causality, not a fully committed thread schedule.

For concurrent scenarios, $\Phi$ is set-valued rather than schedule-singleton. If incomparable events can race, the allowed outcome class for event $e$ may be
\[
\Phi_e = \bigcup_{\pi \in \execset(\mathcal{S},N_A)} \Phi_e^{\pi},
\]
where $\Phi_e^{\pi}$ is the modeled observation class of $e$ under linearization $\pi$. This lets the oracle accept multiple race-permitted outcomes while still rejecting classes that no legal linearization admits.

\section{Scenario Synthesis}
\label{sec:synthesis}

\subsection{Three Scenario Families}
SyncPetri\ synthesizes three classes of scenarios from the same Petri-net model. The classes can be used independently or composed: in Section~\ref{sec:motivation}, a partial-order legal core is followed by a near-legal boundary probe.

These are fully enabled traces selected to reach semantically uncommon markings rather than merely long traces. We assign a heuristic score
\[
\depthscore(\pi) = \alpha U(M_k) + \beta C(\pi) + \gamma X(\pi),
\]
where $\pi$ is a legal trace ending at marking $M_k$, $U(M_k)$ is a marking-novelty term, $C(\pi)$ measures structural coverage (for example, distinct transition or place classes), and $X(\pi)$ measures conflict exposure induced by the trace.

These consist of a legal prefix followed by one near-legal event. They target boundary checks, stale-resource handling, and error propagation without collapsing into meaningless invalid sequences. These begin as legal event sets, but independent steps are left unordered. The output is therefore a DAG of causality plus a conflict relation, not a single linear schedule.

\subsection{Generation Algorithm}
\begin{algorithm}[t]
\caption{Petri-net scenario synthesis}
\label{alg:synth}
\begin{algorithmic}[1]
\Require net $N_A$, family $f$, length bound $L$
\Ensure abstract scenario $\mathcal{S}$
\State $M \gets M_0$, $E \gets [\,]$, $\prec \gets \emptyset$, $\# \gets \emptyset$
\For{$i = 1$ to $L$}
  \State $C_{\text{en}} \gets \{ t \in T \mid \enabled(M,t) \}$
  \State $C_{\text{near}} \gets \{ t \in T \mid |\Delta_F(M,t)| + |\Delta_G(M,t)| = 1 \}$
  \If{$f = \textsc{NearLegal}$ and $i$ is the mutation point}
    \State choose $t^\star$ from $C_{\text{near}}$
    \State append boundary event $e_i$ with $\eta(e_i)=t^\star$
    \State \textbf{break}
  \Else
    \State choose $t^\star$ from $C_{\text{en}}$ maximizing $\depthscore$
    \State append legal event $e_i$ with $\eta(e_i)=t^\star$
    \State add causal edges induced by token production and consumption
    \State add conflict edges induced by shared resources
    \State fire $t^\star$ and update $M$
  \EndIf
\EndFor
\If{$f = \textsc{PartialOrder}$}
  \State remove unnecessary order edges while preserving causality
\EndIf
\State return $\mathcal{S}=(E,\prec,\#,\eta,\beta,\Phi)$
\end{algorithmic}
\end{algorithm}
The Algorithm~\ref{alg:synth} is intentionally model-centric. The choice of which event should happen next is driven by reachability and conflict structure, not by code-generation convenience.

In practice, near-legal mutation is not inserted uniformly. After a legal prefix reaches marking $M$, we score candidate boundary events by
\[
\begin{aligned}
\mathsf{MutScore}(M,t) = {} &
w_1 \mathbf{1}[|\Delta_F(M,t)| + |\Delta_G(M,t)| = 1] \\
& + w_2 \mathsf{Stale}(M,t)
+ w_3 \mathsf{ConflictCtx}(M,t),
\end{aligned}
\]
where $\mathsf{Stale}$ rewards operations that reuse recently invalidated resources and $\mathsf{ConflictCtx}$ rewards mutations placed near a high-conflict frontier. This focuses budget on the kinds of boundary mistakes that concurrent libraries commonly mishandle.

\section{LLM Concretization}
\label{sec:concretization}

\subsection{Prompt Contract}
The LLM receives a typed prompt assembled from the scenario tuple and API adapter schema:
\[
P(\mathcal{S},A_d) = (\mathsf{Hdr}, \mathsf{Res}, \mathsf{Ev}, \mathsf{Ord}, \mathsf{Conf}, \mathsf{Obs}, \mathsf{Out}, \mathsf{Ban}).
\]
The fields encode resource declarations, typed events, order edges, concurrency hints, expected observation classes, and banned behaviors such as invented semantic events or weakened assertions. A representative prompt fragment is shown in Listing~\ref{lst:prompt}.

\begin{lstlisting}[style=ruststyle,caption={Constrained prompt fragment for concretization.},label={lst:prompt}]
scenario_id: mpsc_permit_close
target_api: tokio::sync::mpsc
resources:
  permit_sender: Sp
  boundary_sender: Sb
  receiver: R0
events:
  - e1: reserve_owned Sp -> P0
  - e2: spawn T1 with P0
  - e3: permit_send P0 m1
  - e4: close R0
  - e5: recv R0 -> m1
  - e6: try_send Sb boundary_msg
  - e7: recv_end R0
order:
  - e1 < e2 < e3
  - e1 < e4
  - e3 < e5
  - e4 < e5 < e6 < e7
concurrent:
  - (e3, e4)
expected:
  - class(out(e3)) in {SendOk}
  - class(out(e6)) in {ClosedLike}
  - class(out(e7)) in {End}
output_contract:
  - emit one marker before each modeled event
  - preserve all order constraints
  - return Rust test code only
\end{lstlisting}

The LLM output is represented as $Y = (c, \mu, a)$ where $c$ is Rust test code, $\mu$ maps scenario events to emitted runtime markers, and $a$ is a set of generated assertions.

We accept a generated artifact for execution only if
\[
\begin{aligned}
\mathsf{WellFormed}(Y,\mathcal{S}) \iff {} &
\mathsf{Compiles}(c) \\
& \land \forall e \in E,\; \mu(e)\downarrow \\
& \land \mathsf{NoInventedEvents}(Y,E).
\end{aligned}
\]
This is a static admission filter; the structural oracle later checks whether the runtime trace actually respects the intended partial order.

For evaluation, we also use a partial structural-fidelity score. Let $h^\star$ be a maximum-cardinality order-preserving partial matching from scenario events to emitted markers. We define
\[
\fidelity(\mathcal{S},\tau) = \frac{|{\rm dom}(h^\star)|}{|E|}.
\]
This allows tool to measure how much of the intended structure survives concretization even when the full structural oracle fails. The crucial rule is that the LLM may choose syntax, helper names, and local scaffolding, but it may not invent new semantic events, relax order constraints, or redefine the expected outcome classes.

\subsection{Concretization and Repair Loop}
We require a generated test to compile and to expose the marker structure needed by the runtime oracle. Algorithm~\ref{alg:concretize} gives the loop. This loop is deliberately strict because successful compilation alone is insufficient to prove that the scenario has been correctly concretized.

\begin{algorithm}[t]
\caption{Concretization with structural repair}
\label{alg:concretize}
\begin{algorithmic}[1]
\Require scenario $\mathcal{S}$, adapter schema $A_d$, LLM $L$, retry bound $R$
\Ensure faithful test artifact or failure
\For{$r = 1$ to $R$}
  \State build prompt $P(\mathcal{S}, A_d)$
  \State $Y \gets L(P)$
  \If{$Y.c$ fails to compile}
    \State feed compiler diagnostics back to $L$
    \State \textbf{continue}
  \EndIf
  \If{marker coverage is incomplete}
    \State feed structural diagnostics back to $L$
    \State \textbf{continue}
  \EndIf
  \State return $Y$
\EndFor
\State return \textsc{Fail}
\end{algorithmic}
\end{algorithm}

\section{Petri-Guided Schedule Exploration}
\label{sec:schedule}

The scenario tuple already tells us which events are causally constrained and which pairs are in semantic conflict. We use that information to shape schedule exploration rather than treating all harness variants as equally important.

For a prefix $\rho \subseteq E$, define the ready frontier as
\[
\frontier_{\mathcal{S}}(\rho) = \{ e \in E \setminus \rho \mid \forall e' \prec e,\; e' \in \rho \}.
\]
For any ready event $e$, we define a conflict-first priority
\begin{align*}
\mathsf{prio}_{\rho}(e) = {} &
\alpha \sum_{e' \in \frontier_{\mathcal{S}}(\rho) \setminus \{e\}} \mathbf{1}[(e,e') \in \#] \\
& + \beta \,\mathsf{Rare}(e) - \gamma \,\mathsf{Seen}(\rho,e),
\end{align*}
where $\mathsf{Rare}(e)$ rewards uncommon transition classes and $\mathsf{Seen}$ penalizes already explored prefixes.

A partial-order scenario also induces a task skeleton
\[
K_{\mathcal{S}} = (V_{\text{task}}, E_{\text{spawn}}, E_{\text{conf}}),
\]
where $V_{\text{task}}$ partitions events by task ownership, $E_{\text{spawn}}$ records spawn-parent relations, and
\[
E_{\text{conf}} = \{ (e_i,e_j) \mid e_i \parallel e_j \land (e_i,e_j)\in\# \}
\]
collects incomparable conflict pairs. A concrete harness variant chooses task creation order, barrier placement, and explicit yield points around selected pairs in $E_{\text{conf}}$.

We do not need to modify Loom internals to use this signal. Instead, we generate a small set of \emph{schedule-shaping harness variants} that prioritize different high-conflict frontier choices, then run each variant under Loom when a Loom-compatible harness exists. For APIs where direct Loom integration is unavailable, the same schedule-shaping variants can be executed under a deterministic scheduler wrapper. The improvement is therefore above Loom rather than inside Loom: the Petri net chooses which concurrency skeletons deserve schedule budget.

\begin{algorithm}[t]
\caption{Petri-guided schedule shaping}
\label{alg:loom}
\begin{algorithmic}[1]
\Require partial-order scenario $\mathcal{S}$, variant budget $K$
\Ensure harness variants $H$
\State $H \gets \emptyset$
\For{$j = 1$ to $K$}
  \State $\rho \gets \emptyset$, $h_j \gets [\,]$
  \While{$\rho \neq E$}
    \State $F \gets \frontier_{\mathcal{S}}(\rho)$
    \State choose $e^\star \in F$ maximizing $\mathsf{prio}_{\rho}(e)$
    \State append $e^\star$ to $h_j$
    \State insert yield/barrier hooks around conflicting incomparable pairs
    \State $\rho \gets \rho \cup \{e^\star\}$
  \EndWhile
  \State add $h_j$ to $H$
\EndFor
\State return $H$
\end{algorithmic}
\end{algorithm}

Let $\mathsf{YieldPts}(h)$ denote the conflict pairs that a harness variant $h$ explicitly exposes with barriers or yields, and let $C_{j-1}$ be the set already covered by earlier variants. We can then rank candidate variants by uncovered conflict gain:
\[
h_j =
\arg\max_{h \in \mathcal{H}(\mathcal{S})}
\sum_{(e,e') \in \mathsf{YieldPts}(h)\setminus C_{j-1}} w(e,e').
\]
This gives a concrete methodological improvement over naive schedule enumeration: Loom still explores schedules within each harness, but the Petri layer decides which harnesses deserve schedule budget by maximizing semantically meaningful conflict coverage first.

\section{Multi-Layer Semantic Oracle}
\label{sec:oracle}

The execution harness records a trace
\[
\tau = \big((m_1,o_1), (m_2,o_2), \dots, (m_k,o_k)\big),
\]
where each $m_i$ is an emitted marker and each $o_i$ is the corresponding local observation or return class.

We define the oracle as a conjunction of four layers:
\[
\oracle(\mathcal{S},\tau) =
\oracle_{\text{str}} \wedge
\oracle_{\text{out}} \wedge
\oracle_{\text{inv}} \wedge
\oracle_{\text{live}}.
\]

\paragraph{Structural oracle.}
\[
\oracle_{\text{str}}(\mathcal{S},\tau)=1
\]
iff there exists an injective matching $h : E \rightarrow \{1,\dots,k\}$ such that:
\begin{enumerate}
  \item the marker at position $h(e)$ corresponds to event $e$;
  \item if $e_i \prec e_j$, then $h(e_i) < h(e_j)$.
\end{enumerate}
If $\oracle_{\text{str}}=0$, the test is treated as a concretization failure rather than an API bug.

\paragraph{Outcome oracle.}
Each event $e$ may carry an allowed set of observation classes $\Phi_e$. Then
\[
\oracle_{\text{out}}(\mathcal{S},\tau)=1
\]
iff for every matched event $e$, the observed class at $h(e)$ belongs to $\Phi_e$. This allows the oracle to express sets of admissible concurrent outcomes rather than brittle single-value expectations.

\paragraph{Invariant oracle.}
Let $\Psi$ be the set of global scenario invariants. Then
\[
\oracle_{\text{inv}}(\mathcal{S},\tau)=1
\]
iff every invariant in $\Psi$ holds over the observed execution summary. Typical examples include resource conservation, absence of ghost messages, bounded buffer counts, or monotonic closure state.

\paragraph{Liveness oracle.}
\[
\oracle_{\text{live}}(\mathcal{S},\tau)=1
\]
iff the run terminates within a bound and no task expected to complete remains permanently blocked. This matters because many concurrency faults manifest as hangs rather than wrong return values.

Let
\(
\oracle_{\text{sem}}=
\oracle_{\text{out}}\wedge
\oracle_{\text{inv}}\wedge
\oracle_{\text{live}}
\).
We classify outcomes by
\[
\mathsf{Class}(\mathcal{S},\tau)=
\begin{aligned}
&\textsc{ConcretizationError}
&&\text{if } \oracle_{\text{str}}=0,\\
&\textsc{SemanticFailure}
&&\text{if } \oracle_{\text{str}}=1 \land \oracle_{\text{sem}}=0,\\
&\textsc{Pass}
&&\text{otherwise.}
\end{aligned}
\]
This decision rule is operationally important: only the second case becomes a bug candidate.

Table~\ref{tab:oracle} summarizes the role of each layer.

\begin{table}[t]
\caption{Semantic oracle layers.}
\label{tab:oracle}
\centering
\small
\setlength{\tabcolsep}{4pt}
\begin{tabular}{@{}p{0.18\columnwidth}p{0.25\columnwidth}p{0.47\columnwidth}@{}}
\toprule
Layer & Checks & Failure signal \\
\midrule
Structural & Marker coverage and order preservation & LLM omitted an event or reordered a required edge \\
Outcome & Return/error class membership & A close-after-send boundary event unexpectedly succeeds \\
Invariant & Scenario-level semantic properties & Buffered count becomes inconsistent with sends and receives \\
Liveness & Bounded completion and join behavior & Test hangs after a race that should terminate \\
\bottomrule
\end{tabular}
\end{table}

This separation is important. It prevents the system from confusing a bad generated test with a bad library behavior.

\paragraph{Soundness intuition.}
Assume that (1) the adapter $A_d$ is locally faithful, (2) marker $\mu(e)$ is
emitted immediately before the concrete step for $e$, and (3) helper code emits
no spurious modeled markers. If
\[
\oracle_{\text{str}}(\mathcal{S},\tau)=1,
\]
then $\tau$ contains a matched subsequence with the same order as some
$\pi \in \execset(\mathcal{S},N_A)$. In other words, the runtime trace
refines a modeled linearization of the intended scenario up to unmodeled helper
steps. Under this assumption, a semantic-oracle failure points to API behavior
under the concretized scenario rather than to missing events or reordered
scaffolding.

\paragraph{Failure report.}
When a run fails, the system reports
\[
\mathsf{Report} =
(\mathcal{S}, \mathsf{Class}, L_f, \tau^\star, h_j),
\]
where $L_f \subseteq \{\text{out},\text{inv},\text{live}\}$ is the set of failed oracle layers, $\tau^\star$ is the matched event subsequence, and $h_j$ is the harness variant that exposed the behavior. This report structure matters in practice because it supports triage, de-duplication, and prompt repair without collapsing all failures into a single undifferentiated bug bucket.

\section{Evaluation and Results}
\label{sec:evaluation}
This section reports a preliminary evaluation of the reviewed MPSC running example. The goal is not a large benchmark, but a paper-level check that the current prototype can carry one manually modeled CPN and one reviewed \texttt{ScenarioBlueprint} all the way to executable Tokio code, and that the resulting runtime trace agrees with the model-derived oracle.

\subsection{Subjects and Setup}

The evaluated subject is a bounded Tokio MPSC channel with capacity one. The reviewed scenario contains seven events and exactly two legal linearizations:

\begin{lstlisting}[style=ruststyle]
e3_before_e4: e1 e2 e3 e4 e5 e6 e7
e4_before_e3: e1 e2 e4 e3 e5 e6 e7
\end{lstlisting}

The repaired generated source compiles, executes both legal schedules inside one Tokio test, and emits 14 JSONL records per run. We evaluate the same source three times with the generic artifact-driven evaluator; all three runs pass and the evaluator reports no findings.

\begin{table}[t]
\caption{Current MPSC prototype evidence.}
\label{tab:mpsc-eval}
\centering
\small
\begin{tabular}{p{0.33\columnwidth}p{0.52\columnwidth}}
\toprule
Property & Result \\
\midrule
Compilation & passed \\
Legal schedules & 2 \\
Runtime executions & 3 \\
Records per execution & 14 \\
Evaluator findings & none \\
\bottomrule
\end{tabular}
\end{table}

\subsection{RQ1: Can the artifact-constrained prompt produce executable Tokio code?}

The generated test compiles under the recorded toolchain and runs to completion under the generic evaluator. This is the first requirement for the pipeline: the prompt and generation contract are strong enough to produce a real Tokio test rather than a sketch or pseudo-code fragment.

The same source also carries the scenario labels, the expected observations, and the runtime evidence points that the evaluator consumes. For this MPSC example, those ingredients were sufficient to obtain three successful executions without any MPSC-specific checker in the evaluation crate.

\subsection{RQ2: Does the generated runner preserve schedule-sensitive structure?}

The scenario distinguishes two legal linearizations by the relative order of \texttt{e3} and \texttt{e4}, and the runner preserves that distinction in the runtime trace. Each execution emits the same 14 records, split evenly across the two schedules, and the evaluator confirms that the observed record order matches the artifact oracle.

This matters because the example is not merely ``some test that passes.'' It is a small schedule family whose legal orders remain visible in the final trace. That is the concrete structural property the prototype is meant to preserve.

\subsection{RQ3: Can feedback repair recover from a concrete mismatch without changing the scenario?}

The repaired source used here is the result of a bounded runtime-feedback repair. The repair adjusted the emitted operation labels to the exact Scenario strings, while preserving the same schedules, the same observable behavior, and the same record count. In other words, repair fixed the contract mismatch rather than rewriting the test into a different one.

This is the relevant form of repair for the current prototype: the generated code may change its surface syntax and evidence plumbing, but it should not silently change the modeled scenario or weaken the oracle.

\subsection{RQ4: Is the artifact boundary strict enough to isolate the generic pipeline from the subject-specific model?}

The generic LLM and evaluator layers consume the portable artifact, not handwritten MPSC logic. The subject-specific semantics live in the family crate: the reviewed CPN, the reviewed scenario, and the runtime observation contract. The generic pipeline then compiles the generated source, executes it, and checks the emitted JSONL records against the artifact-derived oracle.

The limitation is scope. This is still a preliminary evaluation on one running example, not a large-scale mutation study or a schedule-budget benchmark. The result supports feasibility, not statistical superiority.

\subsection{Threats to Validity}

The main threat is scale. The current evaluation covers one reviewed MPSC subject and one repaired generated runner. It does not yet include a mutant corpus, a pure prompt baseline, or a schedule-exploration comparison. Even so, the available evidence is enough for an arXiv submission that claims a working prototype and a validated artifact boundary, rather than a completed benchmark.

\section{Related Work}

\paragraph{Model-based testing with Petri nets.}
Petri nets have long been used for formal modeling, reachability reasoning, and test derivation in stateful systems~\cite{murata1989petri,manral2015petritest}. That line of work establishes why token-based models are useful: they express multiplicity, causality, and conflict more naturally than flat state machines. Our setting adds a missing concretization problem. The output must not stop at an abstract trace; it must become a compilable Rust test with tasks, ownership moves, time bounds, and executable assertions. SyncPetri therefore uses the Petri net as a semantic intermediate representation rather than as the whole testing engine.

\paragraph{Stateful API testing and deep-state exploration.}
REST-ler showed that dependency-aware generation is necessary for stateful APIs because naive request composition rarely reaches meaningful deep states~\cite{restler2018}. StateAFL and later stateful greybox work likewise argue that failures often emerge only after the system enters semantically distinct states~\cite{stateafl2021,statefulgreybox2022}. We share that motivation, but the target and abstraction differ. Those systems focus on network-facing or protocol-facing interfaces, whereas SyncPetri\ targets in-process Rust library APIs whose hard behaviors depend on resource ownership, handle invalidation, reserved capabilities, and schedule-sensitive partial orders. Our near-legal scenarios also aim at semantic boundaries defined by the model, rather than at arbitrary invalid inputs.

\paragraph{LLM-based test generation.}
TitanFuzz and Fuzz4All showed that LLMs can generate diverse tests and programs in domains where manual generators are costly~\cite{titanfuzz2022,fuzz4all2023}. CoverUp further showed that external feedback and structure remain important even for strong models~\cite{coverup2024}. SyncPetri\ adopts that lesson but changes the guidance signal. Instead of relying on open-ended prompting or coverage alone, we provide a typed event graph with resource bindings, order constraints, admissible outcome classes, and banned semantic edits, then judge the result with a structural oracle. The contribution is therefore not simply to use an LLM for testing, but to restrict the LLM to code realization while keeping semantic intent outside the model.

\paragraph{Systematic concurrency exploration.}
Loom provides controlled schedule exploration for Rust concurrent code~\cite{loom_docs}. Our work is complementary rather than competitive. Loom explores micro-interleavings inside a harness; SyncPetri\ tries to make that harness semantically meaningful in the first place by extracting race windows and conflict pairs from the resource model. The schedule-shaping component therefore operates above the scheduler: it prioritizes which concurrency skeletons deserve exploration budget, but it does not modify the scheduler's internal search algorithm.

\paragraph{Positioning.}
Across these lines of work, the missing combination is a model-driven route from resource-aware concurrency semantics to executable Rust tests without delegating the semantic oracle to the LLM. SyncPetri\ occupies that space by combining Petri-net scenario synthesis, constrained concretization, schedule-aware harness construction, and layered runtime judgment in one testing workflow.

\section{Conclusion}

This paper presented SyncPetri, a Petri-net-constrained architecture for LLM-based test generation over concurrent stateful Rust APIs. The main thesis is that model structure should control semantics, boundary mutation, and concurrency exposure, while the LLM should only control executable realization. To support that thesis, we formalized the target domain, defined a scenario language, specified concretization and repair loops, proposed Petri-guided schedule shaping, and introduced a multi-layer semantic oracle. Together, these pieces form a main-track-sized methodology: not merely a sketch that Petri nets and LLMs can be combined, but a concrete design for how they should be combined in a test-generation system.

\bibliographystyle{ACM-Reference-Format}
\bibliography{petri_llm_sync_refs}

\end{document}